*Article*

# Deep ensemble learning for segmenting tuberculosis-consistent manifestations in chest radiographs

Sivaramakrishnan Rajaraman[1*], Feng Yang[1], Ghada Zamzmi[1], Peng Guo[1], Zhiyun Xue[1] and Sameer K Antani[1]

[1]National Library of Medicine, National Institutes of Health, Bethesda, MD, USA 20894
*Correspondence: sivaramakrishnan.rajaraman@nih.gov

**Abstract:** Automated segmentation of tuberculosis (TB)-consistent lesions in chest X-rays (CXRs) using deep learning (DL) methods can help reduce radiologist effort, supplement clinical decision-making, and potentially result in improved patient treatment. The majority of works in the literature discuss training automatic segmentation models using coarse bounding box annotations. However, the granularity of the bounding box annotation could result in the inclusion of a considerable fraction of false positives and negatives at the pixel level that may adversely impact overall semantic segmentation performance. This study (i) evaluates the benefits of using fine-grained annotations of TB-consistent lesions and (ii) trains and constructs ensembles of the variants of U-Net models for semantically segmenting TB-consistent lesions in both original and bone-suppressed frontal chest X-rays (CXRs). We evaluated segmentation performance using several ensemble methods such as bitwise AND, bitwise-OR, bitwise-MAX, and stacking. We observed that the stacking ensemble demonstrated superior segmentation performance (Dice score: 0.5743, 95% confidence interval: (0.4055,0.7431)) compared to the individual constituent models and other ensemble methods. To the best of our knowledge, this is the first study to apply ensemble learning to improve fine-grained TB-consistent lesion segmentation performance.

## 1. Introduction

Tuberculosis (TB) is an infectious disease caused by the *Mycobacterium tuberculosis* bacteria. The disease continues to remain the primary cause of ill-health and mortality across the world according to the recent 2020 global report on TB from the World Health Organization (WHO) [1]. Pulmonary infection from the *Mycobacterium* is reported to affect all age groups and sexes. Early screening and diagnosis would therefore prove to be critical to improving chances of survival and patient care. While CT imaging is the preferred diagnostic imaging technique because of its sensitivity in detecting TB, they have several limitations such as reduced access in low and medium resourced regions, high cost, high radiation dose, lack of portability, and increased need for frequent sanitation, among others [2]. Therefore, CXR imaging continues to be the most widely used examination for pulmonary TB-related disease screening [3], particularly in developing countries with limited technical and human resources.

### 1.1. Related literature

Semantic segmentation methods associate each image pixel with a class label. Automatic deep learning (DL)-based algorithms are shown to deliver superior performance in delineating and identifying disease-specific manifestations in CXRs, particularly TB [3–6]. They can help supplement human expertise for clinical decision-making thereby facilitating prompt referrals and subsequently improving patient care. However, the performance of DL algorithms depends on the characteristics of the data under study. A major limitation of current DL algorithms is that they use coarse bounding-box annotations of TB-consistent lesions for training and validation [7]. This might result in including a considerable fraction of false-positive and false-negative pixels in the annotations since the TB infection-specific regions of interest (ROIs) are relatively small and there is variability in the granularity of expert annotations used to train the models. To address this limitation, we propose the first study that uses fine-grained annotations of TB-consistent lesions to train and evaluate machine learning (ML) models.

DL models learn through stochastic backpropagation [8]. Due to the varying architecture and hyperparameters of the model and the stochastic nature of learning, these models may converge to different local optima. Ensemble learning is an established ML paradigm that seeks to improve robustness and accuracy by combining the predictions of several models [9]. Several methods of ensemble learning (e.g., averaging, bagging, boosting, and stacking) are shown to deliver superior performance in medical image segmentation tasks using CXRs. The authors of [10] performed an averaging ensemble of the predictions of the U-Net [11] and DeepLabV3+ models to segment lungs in CXRs. Their proposed method achieved a segmentation accuracy of 98.6% using the Japanese Radiological Scientific Technology (JRST) [12] and Shenzhen CXR [13] datasets. In another study [14], the authors proposed an ensemble DeepLabV3+ based architecture to segment lungs in the Shenzhen CXR collection and achieved an Intersection-Over-Union (IoU) score of 0.97. Ensemble methods were applied to segment pneumothorax-consistent regions [15] in CXRs. The authors used various ImageNet-pretrained encoder backbones in the U-Net model and performed a weighted averaging ensemble of their predictions to segment pneumothorax-consistent regions with a dice score of 0.906.

### 1.2. Contributions of the study

In this study, we propose to (i) use fine-grained annotations of TB-consistent lesions to train variants of U-Net models and (ii) construct an ensemble of the best-performing models to further improve the robustness and performance of the segmentation algorithm. A block diagram that summarizes our study is shown in Figure 1. In the first step, we used a feature pyramid network (FPN)-based model with the EfficientNet-B0 encoder backbone from our previous study [16] to suppress bones in the Shenzhen TB CXR collection. Then, U-Net models with varying ImageNet-pretrained encoder backbones, viz., ResNet-34 [17], Inception-V3 [18], DenseNet-121 [19], EfficientNet-B0 [20], and SE-ResNext-50 [21] were trained and evaluated on the original and bone-suppressed CXRs for segmenting TB-consistent lesions. The predictions of the top-K (K=3, 4, 5) models were used to construct ensemble predictions using several bitwise operations, viz., bitwise-AND, bitwise-OR, and bitwise-MAX. We further constructed a stacking ensemble by concatenating the features extracted from the penultimate layer of the top-K models and training a fully-convolutional meta-learner to optimally combine these features and improve segmentation performance. To the best of our knowledge, this is the first study to (i) use fine-grained annotations of TB-consistent lesions and bone-suppressed CXRs to train and evaluate U-Net models and (ii) improve performance through evaluating several ensemble methods and selecting the best ensemble method for final predictions. We discuss the methods in Section 2, results in Section 3, conclusions, and future work in Section 4.

**2. Materials and Methods**

*2.1. Datasets*

We used the Shenzhen TB CXR [13] collection, which contains 662 de-identified CXRs including 336 TB cases and 326 normal cases. TB cases are either microbiologically confirmed, or with clinical symptoms and imaging appearance consistent with TB, and positive response to anti-TB medication while excluding other causes. The CXRs are collected from patients at the Shenzhen No.3 hospital in Shenzhen, China. The use of these CXRs is exempted from IRB review (OHSRP#5357) by the National Institutes of Health (NIH) and is made publicly available. CXRs manifesting TB-consistent abnormalities are annotated by two radiologists from the Chinese University of Hong Kong using the Firefly annotation tool[1]. The labeling was initially conducted by a junior radiologist, then the labels were all checked by a senior radiologist, with a consensus reached for all cases. The annotations are prepared in JavaScript object notation syntax (JSON) format. They are also prepared as separate grayscale mask images showing abnormal ROIs. The CXRs and their associated masks are resized to 256 × 256 spatial dimensions to reduce computational complexity. The resized CXRs and masks are split into 70% for training, 20% for validation, and 10% for testing. The CXRs used for training are augmented offline using the Augmentor tool [22] using affine transformations including mirroring, rotation in the range [5, 10], and zooming in the range [0.8, 1.4] to create 2000 additional CXRs and their associated masks. The number of CXRs in the train, validation and test sets are shown in Table 1.

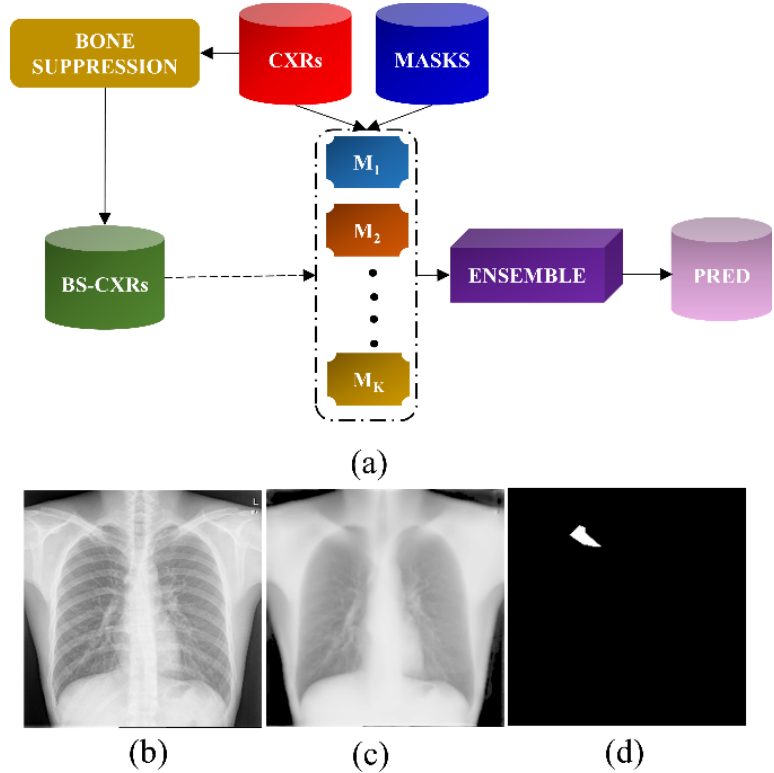

**Figure 1. Block diagram of the proposal**. (a) Process pipeline in training individual U-Net models and constructing ensembles to arrive at the final prediction; (b - d) show a sample CXR, its bone-suppressed counterpart, and the ground truth TB-consistent lesion mask, respectively.

**Table 1. Dataset and its respective patient-level train/validation/test splits**.

| Dataset | Train | Validation | Test |
| --- | --- | --- | --- |
| Shenzhen TB | 2231 | 66 | 33 |

---

[1] https://cell.missouri.edu/software/firefly/

*2.2. Model architecture*

**2.2.1. Bone suppression**

The bones in the Shenzhen TB CXR collection are suppressed using an FPN-based model [23] with an EfficientNet-B0 encoder backbone, which was used in our previous study [16]. The bottom-up pathway extracts image features at multiple scales. The spatial resolution decreases with increasing depth and the semantic value of the layers increases while detecting high-level structures. The top-down pathway constructs high-resolution layers from the semantically rich layers at each scale in the bottom-up pathway. The final layer of the models consists of a convolutional layer with Sigmoidal activation to predict *bone-suppressed* CXRs. The original CXRs and their bone-suppressed counterparts are trained on the augmented NIH-CC-DES-Set 2 dataset and tested with the CXR image pairs in the NIH-CC-DES-Set 1 dataset [16]. The learning rate was reduced whenever no improvement in the validation performance was observed for the subsequent 5 epochs. We used callbacks to store model checkpoints and stopped training when the performance on the validation set began to degrade. The models were trained on an Ubuntu Linux system with NVIDIA GeForce GTX 1080 Ti graphics card using the Keras framework with Tensorflow backend and CUDA dependencies for accelerating the GPUs.

**2.2.2. Segmentation of TB-consistent lesions**

We trained the U-Net variants with ImageNet-pretrained encoder backbones, viz., ResNet-34, Inception-V3, DenseNet-121, EfficientNet-B0, and SE-ResNext-50 to segment the TB-consistent lesions in the CXRs. U-Net has a U-shaped architecture with an encoder followed by a decoder network. The various ImageNet-pretrained models, aforementioned, used in the encoder/contracting path encode the input CXRs into feature representations at multiple scales. The number of feature channels gets doubled with each down-sampling step. The decoder/expanding path up-samples the feature maps to project the low-resolution features into the high-resolution pixel space. The skip connections/concatenations ensure that the low-level information is shared between the input and output, thereby adding information that might be lost because of the down-sampling on the encoder side of the network. The final convolutional layer in the decoder network with Sigmoidal activation predicts the masks. The U-Net models are trained on the augmented Shenzhen TB CXRs and their corresponding TB-consistent lesion masks (from Table 1) using an Adam optimizer with an initial learning rate of $1 \times 10^{-4}$. Callbacks are used to store model checkpoints. The learning rate is reduced whenever the validation performance ceased to improve. The best-performing checkpoint with the validation data is used to predict the test data and generate masks.

**2.2.3. Ensemble learning**

We combined the predictions of the top-K (K = 3, 4, 5) models using the following ensemble methods: (i) bitwise-AND, (ii) bitwise-OR, (iii) bitwise-MAX, and (iv) Stacking, as illustrated in Figure 2. For bitwise operations, we performed a pixel-wise comparison of the predicted masks by the constituent models to construct the final prediction. For a bitwise-AND ensemble, the pixel in the final prediction is turned on only if the corresponding pixels in the predictions from the top-K models are greater than 0. As for a bitwise-OR ensemble, the pixel in the final prediction is turned on if even one of the corresponding pixels in the predictions from the top-K models is greater than 0. We computed the bitwise-MAX across the masks predicted by the top-K models to turn on the corresponding pixel in the final ensemble prediction, otherwise, the pixels are set to 0. We further constructed a stacking ensemble as follows: (i) The features from the penultimate layer of the top-K models are extracted and concatenated; (ii) A fully-convolutional meta-learner performs *second-level* learning to optimally combine these features and improve segmentation performance compared to the constituent models. The meta-learner consists of five convolutional layers. The number of filters in the 1st, 2nd, 3rd, 4th, and 5th convolutional layers are 256, 128, 64, 32, and 1, respectively. All convolutional filters except for the final convolutional layer are of 3 × 3 dimensions and use ReLU activation. The feature maps are equally padded to ensure the output has an identical shape to the input. The trainable weights of the constituent models are frozen and only the fully-convolutional meta-leaner is trained on the extracted features. The final convolutional layer with sigmoidal activation and 1 filter of dimension 1 × 1 predicts the masks. The predicted masks are compared to the ground-truth (GT) masks to evaluate segmentation performance.

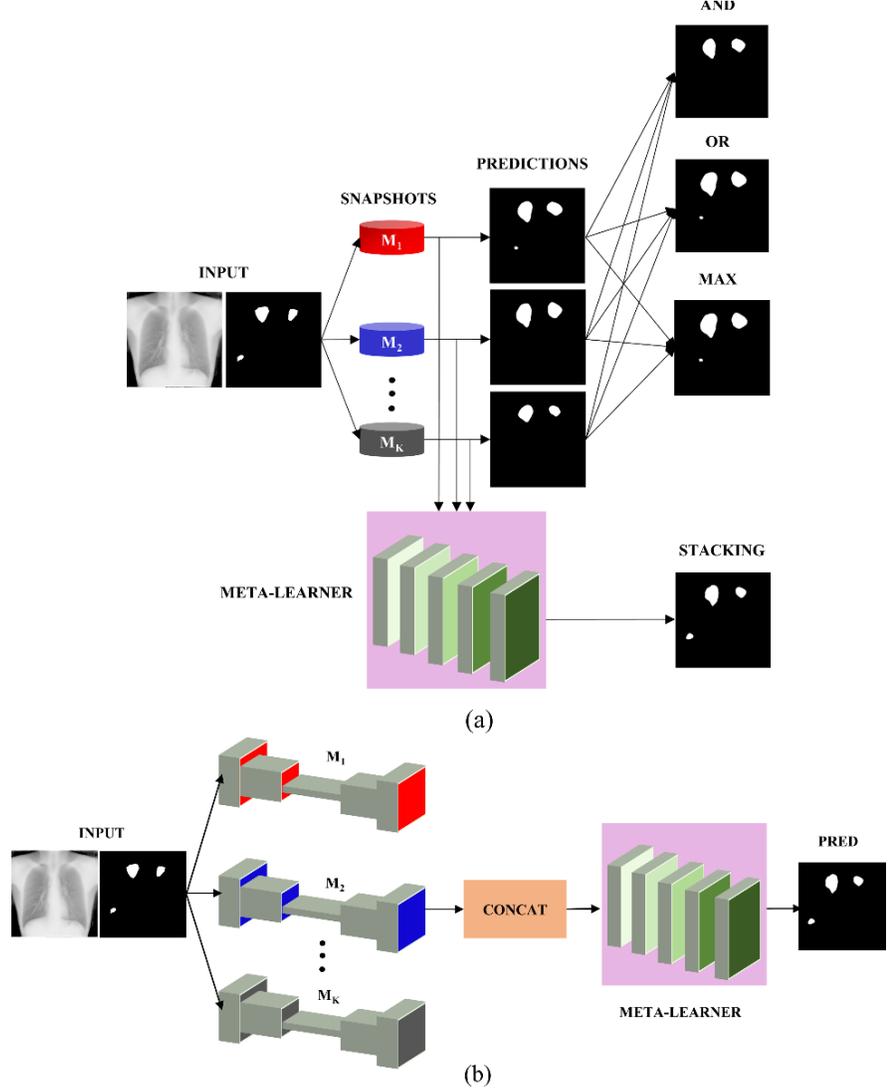

**Figure 2. Ensemble strategies using the predictions of top-K (K = 3, 4, 5) models.** (a) shows the method flowchart showing bitwise-OR, bitwise-AND, bitwise-MAX, and stacking ensemble outputs and (b) shows the method flowchart detailing the stacking process.

### 2.2.4. Loss functions and evaluation metrics

We used an Adam optimizer with an initial learning rate of $1 \times 10^{-3}$ to reduce a mixed loss function given by Equation [1] to train the bone suppression model. Here, MS-SSIM denotes the multi-scale structural similarity index measure and MAE denotes the mean absolute error. The value of $\alpha$ = 0.84 and $\beta$ = 0.16 was empirically determined to deliver superior performance.

$$Mixed - loss = \alpha.MS - SSIM + \beta.MAE \tag{1}$$

The U-Net-based segmentation models are evaluated in terms of Intersection over Union (*IoU*)/Jaccard, Dice/F1-score, and mean average precision (mAP). The *IoU* metric is widely used in evaluating semantic segmentation tasks. It is given by Equations [2] and [3].

$$IoU = TP/(TP + FP + FN) \tag{2}$$

$$IoU_{loss} = 1 - IoU \tag{3}$$

Here, TP, FP, and FN denote the true positives, false positives, and false negatives, respectively, in segmenting the TB-consistent lesions. For a given *IoU* threshold, a predicted mask is considered to be TP if it overlaps with the GT mask by a value exceeding this threshold. FP denotes that the predicted mask has no associated GT mask. FN denotes the GT mask has no associated predicted mask.

The Dice score is another widely used segmentation evaluation metric, given by Equations [4] and [5]. The value of the Dice score ranges from [0, 1] like the *IoU* metric. Higher values of the IoU and Dice score denote improved similarity between the predicted and GT masks.

$$Dice\ score = 2 * TP/(2 * TP + FP + FN) \tag{4}$$

$$Dice_{loss} = 1 - Dice\ score \tag{5}$$

The mean average precision (mAP) is measured as the area under an 11-point interpolated precision-recall curve (AUPRC), given by Equations [6] – [8]. Here, Precision measures the accuracy of predictions, and Recall measures how well the model identifies all the TPs. The Recall values are segmented evenly into 11 parts, i.e., {0, 0.1, 0.2, 0.3, ., 0.9, 1.0} and the mAP is calculated by measuring the AUPRC. We fix the IoU threshold as 0.5. The value of mAP lies in the range [0, 1].

$$Precision(P) = \frac{TP}{TP + FP} \tag{6}$$

$$Recall = \frac{TP}{(TP + FN)} \tag{7}$$

$$mAP = \frac{1}{11} \sum_{Recall_i} Precision(Recall_i) \tag{8}$$

For segmenting the TB-consistent lesions and to enforce all models to have a high recall, we included the boundary uncertainty (BU) evaluation [24] while minimizing the Focal Tversky (FT) loss function [25], given by Equation [9]. The FT loss is parameterized by $\gamma$ to balance between the majority background and minority TB-consistent lesion (ROI) pixels. The value *TI*, given by Equation [10], denotes the Tversky index (TI) function [25] which generalizes the Dice score. Here, $c$ denotes the minority TB-consistent ROI class.

$$FT_{loss}(p,p')_c = \sum_c 1 - TI_c^\gamma \tag{9}$$

$$TI(p,p') = \frac{pp'}{pp' + \lambda(1-p)p' + (1-\lambda)p(1-p')} \tag{10}$$

The value of $\lambda$ lies in the range [0, 1]. When $\lambda$ = 0.5, the equation simplifies to the regular Dice score. Higher values of $\lambda$ will penalize the FNs more than the FPs. That is, with higher values of $\lambda$, the FNs will be kept low with increasing recall since we are concerned with how well the model identifies all the TPs. After empirical evaluations, we fixed the value of $\lambda = 0.7$ and $\gamma = 0.75$.

In a binary segmentation problem, each pixel $t$ in the GT mask, at location $x$, is assigned a hard class label as shown in Equation [11].

$$t_x : \begin{cases} t = 1, if\ x \in \mathbb{F} \\ t = 0, if\ x \notin \mathbb{F} \end{cases} \quad (11)$$

Here, $\mathbb{F}$ denotes the target. While evaluating BU, the hard labels 0 and 1 are converted into soft labels to represent probabilistic scores.

$$t_x : \begin{cases} t \leq 1, if\ x \in \mathbb{F} \\ t \geq 0, if\ x \notin \mathbb{F} \end{cases} \quad (12)$$

$$t_{x \notin \mathbb{F}} \leq t_{x \in \mathbb{F}} \quad (13)$$

Equations [12] and [13] explain that the values closer to 1 and 0 denote higher confidence in classifying the pixels as belonging to the TB disease-consistent ROI or background respectively. The soft labels are restricted only to the ROI boundaries to approximate the uncertainty in segmentation using morphological operators such as dilation ($\triangle$) and erosion ($\triangle$) [24]. Let $X$ denote the input image of dimension $a \times b$. The BU function performs dilation and erosion operations on the ROI boundaries at all positions by querying with a structural element $Y$ of $3 \times 3$ spatial dimensions as shown in Equations [14] and [15]. Probabilities are then assigned for the pixels on the ROI boundaries as shown in Equation [16].

$$(X \triangle Y)(x,y) = max_{i \in S1 j \in S2}\ (X(x - i, y - j) + Y(i,j)) \quad (14)$$

$$(X \triangle Y)(x,y) = min_{i \in S1 j \in S2}\ (X(x + i, y + j) - Y(i,j)) \quad (15)$$

$$t_{x \in \mathbb{F}} : \begin{cases} t = \zeta, if\ t\ \in ((X \triangle Y)_n - X) \\ t = \Omega, if\ t\ \in (X - (X \triangle Y)_n) \end{cases} \quad (16)$$

Here, $n = 1$ denotes the number of iterations for which the erosion and dilation operations are performed. The hyperparameters $\zeta$ and $\Omega$ denote the values for the soft labels that are exterior and interior to the ROI boundaries respectively. When $\zeta = 1$ and $\Omega = 0$, the soft labels would converge to the original hard labels. After empirical evaluations, we fixed the value of $\zeta = 0.9$ and $\Omega = 0.1$. This BU component is incorporated with the FT loss function to train the U-Net variants toward segmenting TB-consistent lesions.

### 2.2.5. Statistical analysis

We evaluated the statistical significance of the reported Dice score with the hold-out test data. We measured the 95% confidence intervals (CIs) as the binomial Clopper-Pearson interval for the reported Dice scores. We followed the guidelines in [26] to measure the $p$-values from the CIs.

## 3. Results and discussions

Recall that the U-Net models are trained on the original and bone-suppressed CXRs. Table 2 shows their TB-consistent ROI segmentation performance. Figure 3 shows the receiver-operating-characteristic (ROC) curves, precision-recall (PR) curves, and confusion matrices obtained using the top-performing models trained on original and bone-suppressed CXRs, respectively.

**Table 2. TB-consistent lesion segmentation performance using original (O) and bone-suppressed (BS) CXRs**. Values in parenthesis denote the 95% CIs for the Dice score. Bold numerical values denote superior performance.

| Models | IOU | Dice |
| --- | --- | --- |
| ResNet-34 (O) | 0.3599 | 0.5293 (0.3589,0.6997) |
| ResNet-34 (BS) | 0.3280 | 0.4640 (0.2938,0.6342) |
| Inception-V3 (O) | **0.3896** | **0.5608 (0.3914,0.7302)** |
| Inception-V3 (BS) | 0.2525 | 0.4032 (0.2358,0.5706) |
| DenseNet-121 (O) | 0.2996 | 0.4611 (0.2910,0.6312) |
| DenseNet-121 (BS) | 0.2892 | 0.4486 (0.2789,0.6183) |
| EfficientNet-B0 (O) | 0.3453 | 0.5134 (0.3428,0.6840) |
| EfficientNet-B0 (BS) | 0.3381 | 0.5053 (0.3347,0.6759) |
| SE-ResNext-50 (O) | 0.3201 | 0.4850 (0.3144,0.6556) |
| SE-ResNext-50 (BS) | 0.2962 | 0.4570 (0.2870,0.6270) |

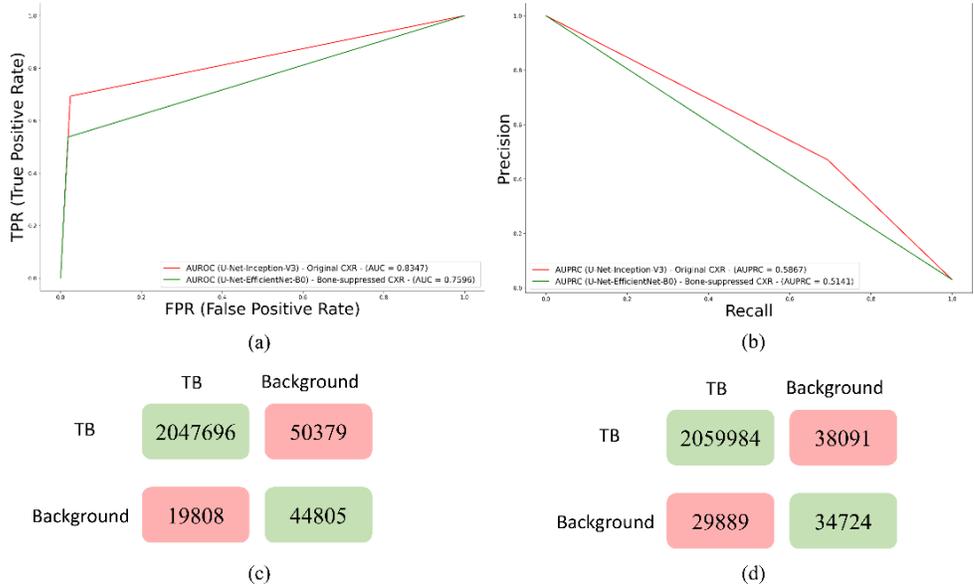

**Figure 3. Performance of the top-performing U-Net with Inception-V3 encoder backbone trained on original CXRs and EfficientNet-B0 encoder backbone trained using bone-suppressed CXRs**. (a) ROC curves; (b) PR curves; (c) Confusion matrix achieved with original CXRs, and (d) Confusion matrix achieved with bone-suppressed CXRs.

It is observed from Table 2 and Figure 3 that the U-Net models trained on the original CXRs outperformed those trained on bone-suppressed CXRs in terms of IoU, Dice score, AUPRC, and area under the receiver-operating characteristic (AUROC) curve. This difference in performance is surprising and counter-intuitive since bone suppression would improve soft-tissue visibility in the image. However, we believe that it can be attributed to the following: (i) The FPN model with the EfficientNet-B0-based encoder was trained and evaluated on a different, sparse, bone-suppressed CXR collection from [16]. The model was trained in-house and not widely tested with cross-institutional datasets. This could have impacted the model's generalization to other datasets due to the heterogeneities in X-ray acquisition and imaging protocols, variability in the overlying cardiopulmonary structures, and bone individualities such as previous fractures and other support devices in the CXR collections. The lack of generalization could result in uneven suppression, brightness, and contrast changes, and irrelevant suppression of the soft tissues, which may adversely impact the detection of apical, central, and basal lesions, and subsequent disease-specific segmentation performance as with TB. With the increased availability of dual-energy subtraction (DES) CXRs, bone suppressed images from the device could be used in the training process which could introduce sufficient data diversity, and therefore help train deeper model architectures to generalize to real-world data. Studies in the literature report that the soft tissue projections obtained with DES systems are superior in quality compared to those generated by DL-based bone suppression methods [27]. Automated bone suppression methods are also shown to be lacking in preserving the frequency details in the original images, therefore small lesions may fade out and go unnoticed upon removing the overlying bony structures in the CXRs [27]. Such phenomena can be noticed in Figure 4. Here, unlike the top-performing U-Net with Inception-V3 encoder backbone trained on original CXRs, the top-performing U-Net model with EfficientNet-B0 encoder backbone trained using bone-suppressed CXRs failed to segment the TB-consistent lesion. These factors might have contributed to the reduction in segmentation performance using the bone-suppressed CXRs. Therefore, we used the models trained on original CXRs to construct ensemble predictions to further improve segmentation performance.

The TB-consistent lesion segmentation performance achieved with various ensemble methods is compared to the best-performing U-Net model with the Inception-V3 encoder backbone (baseline) as shown in Table 3. Figure 5 shows the ROC curves, PR curves, and confusion matrices obtained using the U-Net with Inception-V3 encoder backbone and the stacking ensemble constructed using the top-3 performing models trained on original CXRs. The predicted masks using these models for a sample CXR are shown in Figure 6.

**Table 3. TB-consistent lesion segmentation performance using various ensemble methods. The ensemble performance is compared to the top-performing U-Net model with the Inception-V3 encoder backbone (baseline)**. Values in parenthesis denote the 95% CIs for the Dice score. Bold numerical values denote superior performance.

| Models | IOU | Dice |
| --- | --- | --- |
| Inception-V3 (O) | 0.3896 | 0.5608 (0.3914,0.7302) |
| Top-3 ensemble | | |
| Stacking | **0.4028** | **0.5743 (0.4055,0.7431)** |
| Bitwise-AND | 0.3829 | 0.5538 (0.3841,0.7235) |
| Bitwise-OR | 0.3558 | 0.5249 (0.3545,0.6953) |
| Bitwise-MAX | 0.3343 | 0.5011 (0.3305,0.6717) |
| Top-4 ensemble | | |
| Stacking | 0.3962 | 0.5675 (0.3984,0.7366) |
| Bitwise-AND | 0.3534 | 0.5222 (0.3517,0.6927) |
| Bitwise-OR | 0.3088 | 0.4718 (0.3014,0.6422) |
| Bitwise-MAX | 0.2971 | 0.4581 (0.2881,0.6281) |
| Top-5 ensemble | | |
| Stacking | 0.3974 | 0.5687 (0.3997,0.7377) |
| Bitwise-AND | 0.3534 | 0.5222 (0.3517,0.6927) |
| Bitwise-OR | 0.3088 | 0.4718 (0.3014,0.6422) |
| Bitwise-MAX | 0.2744 | 0.4306 (0.2616,0.5996) |

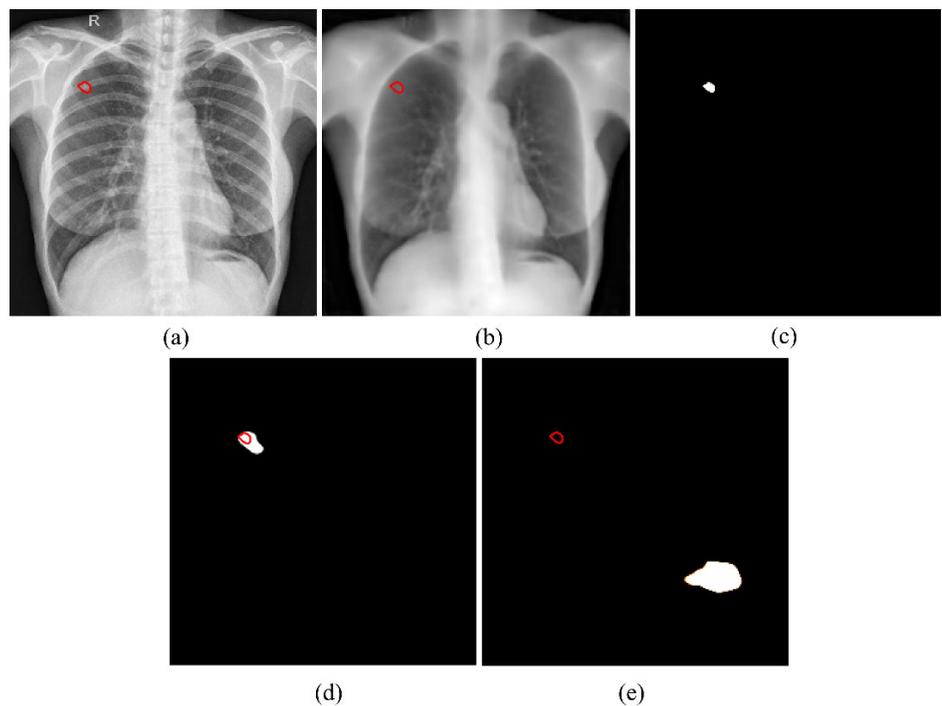

(a)　　　　　　(b)　　　　　　(c)

(d)　　　　　　(e)

**Figure 4. TB-consistent lesion segmentation performance using the U-Net with Inception-V3 encoder backbone trained on original CXRs and EfficientNet-B0 encoder backbone trained using bone-suppressed CXRs**. (a) Sample original CXR image with TB-consistent lesion annotated in red; (b) Corresponding bone-suppressed CXR image with TB-consistent lesion annotated in red; (c) GT TB-consistent lesion mask; (d) Predicted mask using the Inception-V3 encoder backbone trained on original CXRs, and (e) Predicted mask using the EfficientNet-B0 encoder backbone trained using bone-suppressed CXRs.

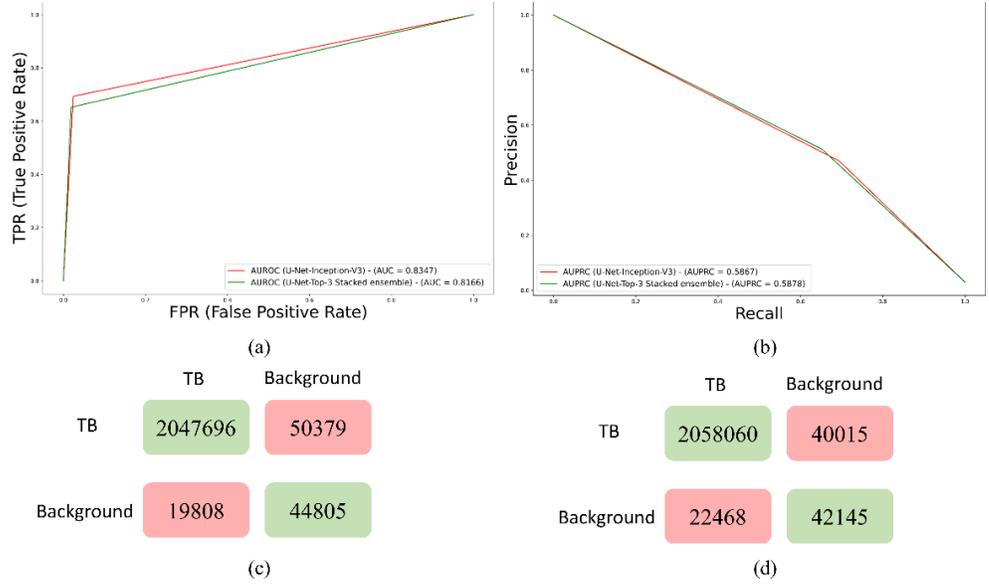

**Figure 5. Performance of the U-Net with Inception-V3 encoder backbone and the stacking ensemble constructed using the top-3 performing models trained on original CXRs.** (a) ROC curves; (b) PR curves; (c) Confusion matrix achieved with Inception-V3 encoder backbone, and (d) Confusion matrix achieved with the stacking ensemble.

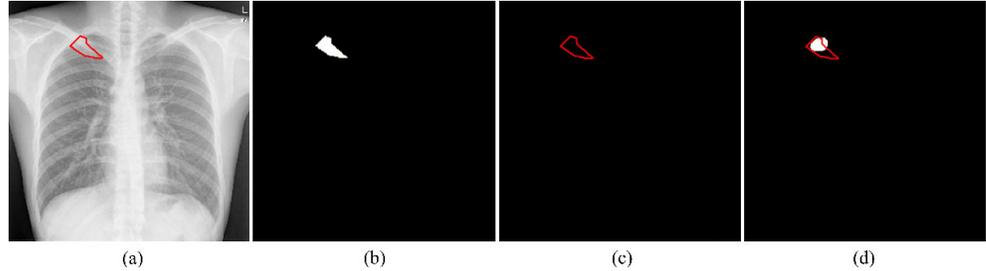

**Figure 6. TB-consistent lesion segmentation performance using the U-Net with Inception-V3 encoder backbone and the stacking ensemble of the top-3 performing models.** (a) Sample original CXR image with TB-consistent lesion annotated in red; (b) GT TB-consistent lesion mask; (c) Predicted mask using the Inception-V3 encoder backbone, and (d) Predicted mask using the stacking ensemble.

The ensemble predictions are generated using the top-K (K = 3, 4, 5) models. We observed that the stacking ensemble using the top-3 performing models, viz. the U-Net model with Inception-V3, ResNet-34, and EfficientNet-B0 encoder backbones, respectively, demonstrated superior segmentation performance in terms of IoU, Dice score, and AUPRC, compared to other ensemble methods and the best-performing U-Net model with the Inception-V3 encoder backbone. The TPs obtained with the stacking ensemble, i.e., the number of TB-consistent lesion pixels segmented correctly is higher than that achieved using the individual top-performing U-Net model with the Inception-V3 encoder backbone model. The improvement in performance using a stacking ensemble could be attributed to the fact that it uses a second-level meta-learner that learns to optimally combine the features learned by the *heterogeneous* base learners, having different architectures, and learning diversified regions in the feature space to converge to their local optima, to output the final prediction. We observed that the stacking ensemble of the top-3 performing models significantly outperformed ($p < 0.05$) the bitwise-OR and bitwise-MAX ensembles constructed using the top-K (K=3, 4, 5) models in terms of the Dice score.

### 4. Conclusion and future work

To the best of our knowledge, this is the first study that uses fine-grained annotations and demonstrates the efficacy of various ensemble methods toward segmenting TB-consistent lesions in both original and bone-suppressed CXRs. This study, however, suffers from the following limitations: (i) CXRs with TB-consistent lesions that are used to train and evaluate the segmentation models are limited. Additional diversity in the training process could be introduced by using CXR data and annotations from cross-institutions. (ii) We used affine transformations to augment the data used for model training. Selecting an appropriate data augmentation technique is a challenging task and it depends on the characteristics of the data under study. Further, data augmentation could introduce more bias into model training if the original dataset contains the same. Therefore, it is critical to empirically identify an appropriate data augmentation strategy such that it creates data variability that can improve the ability of the models to optimally fit the training data while also generalizing to real-world data. Future research could explore other advanced data augmentation methods including adversarial training, neural style transfer, and reinforcement learning, among others. (iii) We used the widely adopted U-Net architecture to segment TB-consistent lesions. Other advanced architectures including FPN, Link-Net, PSP-Net, Bi-SegNet [28], and Trilateral Attention Net [29], among others, and their ensembles could be trained for potential improvements in segmentation performance. With the advent of high-performance computing and storage solutions, the ensemble models could be trained and deployed in the cloud to be used for real-time applications. The methods discussed in this study could be extended to a variety of medical image segmentation tasks.

**Funding:** This research was supported by the Intramural Research Program of the National Library of Medicine, National Institutes of Health.

**Institutional Review Board Statement:** Ethical review and approval were waived for this study because of the retrospective nature of the study and the use of anonymized patient data.

**Informed Consent Statement:** Patient consent was waived by the IRBs because of the retrospective nature of this investigation and the use of anonymized patient data.

**Conflicts of Interest:** The authors declare no conflict of interest.

**References**

1. WHO *World Health Organization Global Tuberculosis Report*;
2. Mettler, F.A.; Huda, W.; Yoshizumi, T.T.; Mahesh, M. Effective Doses in Radiology and Diagnostic Nuclear Medicine: A Catalog. *Radiology* **2008**, *248*, 254–263, doi:10.1148/radiol.2481071451.
3. Jaeger, S.; Karargyris, A.; Candemir, S.; Siegelman, J.; Folio, L.; Antani, S.; Thoma, G. Automatic Screening for Tuberculosis in Chest Radiographs: A Survey. *Quant. Imaging Med. Surg.* **2013**, *3*, 89–99, doi:10.3978/j.issn.2223-4292.2013.04.03.
4. Zamzmi, G.; Rajaraman, S.; Antani, S. UMS-Rep: Unified Modality-Specific Representation for Efficient Medical Image Analysis. *Informatics Med. Unlocked* **2021**, *24*, 100571, doi:10.1016/j.imu.2021.100571.
5. Tang, P.; Yang, P.; Nie, D.; Wu, X.; Zhou, J.; Wang, Y. Unified Medical Image Segmentation by Learning from Uncertainty in an End-to-End Manner. *Knowledge-Based Syst.* **2022**, *241*, doi:10.1016/j.knosys.2022.108215.
6. Rajaraman, S.; Folio, L.R.; Dimperio, J.; Alderson, P.O.; Antani, S.K. Improved Semantic Segmentation of Tuberculosis—Consistent Findings in Chest x-Rays Using Augmented Training of Modality-Specific u-Net Models with Weak Localizations. *Diagnostics* **2021**, *11*, doi:10.3390/diagnostics11040616.
7. Liu, Y.; Wu, Y.H.; Ban, Y.; Wang, H.; Cheng, M.M. Rethinking Computer-Aided Tuberculosis Diagnosis. In Proceedings of the Proceedings of the IEEE Computer Society Conference on Computer Vision and Pattern Recognition; 2020.
8. LeCun, Y.; Bengio, Y. Convolutional Networks for Images, Speech, and Time Series. *Handb. brain theory neural networks* **1995**, *3361*, 255–258, doi:10.1109/IJCNN.2004.1381049.
9. Smolyakov, V. Ensemble Learning to Improve Machine Learning Results.

10. Narayanan, B.N.; De Silva, M.S.; Hardie, R.C.; Ali, R. Ensemble Method of Lung Segmentation in Chest Radiographs. *Proc. IEEE Natl. Aerosp. Electron. Conf. NAECON* **2021**, *2021-Augus*, 382–385, doi:10.1109/NAECON49338.2021.9696439.

11. Ronneberger, O.; Fischer, P.; Brox, T. U-Net: Convolutional Networks for Biomedical Image Segmentation. In Proceedings of the Lecture Notes in Computer Science (including subseries Lecture Notes in Artificial Intelligence and Lecture Notes in Bioinformatics); 2015.

12. Candemir, S.; Jaeger, S.; Antani, S.; Bagci, U.; Folio, L.R.; Xu, Z.; Thoma, G. Atlas-Based Rib-Bone Detection in Chest X-Rays. *Comput. Med. Imaging Graph.* **2016**, doi:10.1016/j.compmedimag.2016.04.002.

13. Jaeger, S.; Candemir, S.; Antani, S.; Wang, Y.-X.J.; Lu, P.-X.; Thoma, G. Two Public Chest X-Ray Datasets for Computer-Aided Screening of Pulmonary Diseases. *Quant. Imaging Med. Surg.* **2014**, *4*, 475–477, doi:10.3978/j.issn.2223-4292.2014.11.20.

14. Ali, R.; Hardie, R.C.; Ragb, H.K. Ensemble Lung Segmentation System Using Deep Neural Networks. *Proc. - Appl. Imag. Pattern Recognit. Work.* **2020**, *2020-Octob*, 0–4, doi:10.1109/AIPR50011.2020.9425311.

15. Abedalla, A.; Abdullah, M.; Al-Ayyoub, M.; Benkhelifa, E. Chest X-Ray Pneumothorax Segmentation Using U-Net with EfficientNet and ResNet Architectures. *PeerJ Comput. Sci.* **2021**, *7*, 1–36, doi:10.7717/peerj-cs.607.

16. Rajaraman, S.; Cohen, G.; Spear, L.; Folio, L.; Antani, S. DeBoNet: A Deep Bone Suppression Model Ensemble to Improve Disease Detection in Chest Radiographs. *PLoS One* **2022**, *17*, 1–22, doi:10.1371/journal.pone.0265691.

17. He, K.; Zhang, X.; Ren, S.; Sun, J. Deep Residual Learning for Image Recognition. In Proceedings of the 2016 IEEE Conference on Computer Vision and Pattern Recognition (CVPR); 2016; Vol. 7, pp. 770–778.

18. Szegedy, C.; Vanhoucke, V.; Ioffe, S.; Shlens, J.; Wojna, Z. Rethinking the Inception Architecture for Computer Vision. *Proc. IEEE Comput. Soc. Conf. Comput. Vis. Pattern Recognit.* **2016**, 2818–2826, doi:10.1002/2014GB005021.

19. Huang, G.; Liu, Z.; Van Der Maaten, L.; Weinberger, K.Q. Densely Connected Convolutional Networks. In Proceedings of the Proceedings - 30th IEEE Conference on Computer Vision and Pattern Recognition, CVPR 2017; 2017.

20. Tan, M.; Le, Q. V. EfficientNet: Rethinking Model Scaling for Convolutional Neural Networks. In Proceedings of the 36th International Conference on Machine Learning, ICML 2019; 2019.

21. Hu, J.; Shen, L.; Sun, G. Squeeze-and-Excitation Networks. *Proc. IEEE Comput. Soc. Conf. Comput. Vis. Pattern Recognit.* **2018**, 7132–7141, doi:10.1109/CVPR.2018.00745.

22. Bloice, M.D.; Roth, P.M.; Holzinger, A. Biomedical Image Augmentation Using Augmentor. *Bioinformatics* **2019**, *35*, 4522–4524, doi:10.1093/bioinformatics/btz259.

23. Xie, X.; Liao, Q.; Ma, L.; Jin, X. Gated Feature Pyramid Network for Object Detection. *Lect. Notes Comput. Sci. (including Subser. Lect. Notes Artif. Intell. Lect. Notes Bioinformatics)* **2018**, *11259 LNCS*, 199–208, doi:10.1007/978-3-030-03341-5_17.

24. Yeung, M.; Yang, G.; Sala, E.; Schönlieb, C.-B.; Rundo, L. Incorporating Boundary Uncertainty into Loss Functions for Biomedical Image Segmentation. **2021**.

25. Abraham, N.; Khan, N.M. A Novel Focal Tversky Loss Function with Improved Attention U-Net for Lesion Segmentation. In Proceedings of the Proceedings - International Symposium on Biomedical Imaging; 2019.

26. Altman, D.G.; Bland, J.M. Statistics Notes: How to Obtain the P Value from a Confidence Interval. *BMJ* **2011**, *343*, 1–2, doi:10.1136/bmj.d2304.

27. Bae, K.; Oh, D.Y.; Yun, I.D.; Jeon, K.N. Bone Suppression on Chest Radiographs for Pulmonary Nodule Detection: Comparison between a Generative Adversarial Network and Dual-Energy Subtraction. *Korean J. Radiol.* **2022**, *23*, 139–149, doi:10.3348/kjr.2021.0146.

28. Badrinarayanan, V.; Kendall, A.; Cipolla, R. Segnet: A Deep Convolutional Encoder-Decoder Architecture for Image Segmentation. *IEEE Trans. Pattern Anal. Mach. Intell.* **2017**, *39*, 2481–2495.

29. Zamzmi, G.; Rajaraman, S.; Sachdev, V.; Antani, S. Trilateral Attention Network for Real-Time Cardiac Region Segmentation. *IEEE Access* **2021**, *9*, 118205–118214, doi:10.1109/ACCESS.2021.3107303.